\documentclass[12pt]{article}
\usepackage{amssymb}
\usepackage{amsfonts}
\def\Rn{{\Bbb R}^n}
\def\Tr{\mathop{\rm Tr}\nolimits}

\begin{document}
\begin{titlepage}
\hbox to \hsize{\hfil IC/98/201} 
\hbox to \hsize{\hfil November, 1998} 
\vfill
\begin{center}
%\title
{ \Large \bf Bering's proposal for boundary contribution to the
Poisson bracket}
\end{center}
\vskip 0.3cm
\begin{center}
%\author
{\bf \normalsize Vladimir O. Soloviev\footnote{e-mail:
soloviev@th1.ihep.su}\\ {\it\small Institute for High Energy
Physics,}\\ {\it\small 142284, Protvino, Moscow region, Russian
Federation}\\ {\it\small and}\\
 {\it\small The Abdus Salam International Center for Theoretical
Physics,}\\ {\it\small P.O. Box 586, 34100 Trieste, Italy}}
\date{}
\end{center}
\vskip 1cm
%\maketitle
\begin{abstract}
It is shown that the Poisson bracket with boundary terms recently
proposed by Bering (hep-th/9806249) can be deduced from the
Poisson bracket proposed
 by the present author
(hep-th/9305133) if one omits  terms free of  Euler-Lagrange
derivatives ("annihilation principle"). This corresponds to
another definition of the formal product of distributions
 (or, saying it in other words, to another definition of the
pairing between 1-forms and 1-vectors in the formal variational
calculus). We extend the  formula (initially suggested by Bering
only for the ultralocal case with constant coefficients) onto the
general non-ultralocal brackets with coefficients depending on
fields and their spatial derivatives. The lack of invariance under
changes of dependent variables (field redefinitions) seems a
drawback of this proposal.
\end{abstract}
\vfill
\end{titlepage}
\section{Introduction}
Recently Bering \cite{Bering} proposed a formula for the field
theory Poisson bracket with boundary terms which are different
from those proposed earlier by the present author \cite{Sol93}. In
general the motivation for new brackets arises from the fact that
the well-known standard field theory Poisson bracket is not
applicable to many nontrivial boundary problems. In particular it
does not satisfy the Jacobi identity. The terms violating the
Jacobi identity are, of course, of purely boundary (or divergence)
nature and so can be killed by some
 boundary conditions which are called trivial here. The problem addressed in this paper as well as in
publications \cite{Bering, Sol93, Sol96, Sol98} consists in
searching for the Poisson bracket formula which exactly fulfils
the Jacobi identity even before any boundary conditions are taken
into account.

According to our knowledge the first observation of this problem and the
first successful attempt to solve it was made by Lewis, Marsden,
Montgomery and Ratiu (LMMR) in their treatment of dynamics of the ideal fluid with a free boundary \cite{lmmr}. Both formulae proposed in \cite{Bering} and
\cite{Sol93} are in fact only  two different extrapolations of the same formula
suggested in \cite{lmmr}. So, it would be better to remind first the
pioneering approach.

In contrast to the popular view \cite{rt} that in the Hamiltonian approach all functionals should be ``differentiable'' the variations of functionals studied by LMMR are not free of
boundary terms
\begin{equation}
D_{q}F(q,p)\cdot \delta
q=\int\limits_{{\Omega}}{{\delta^{\wedge}F}\over{\delta q}} \cdot \delta
q\ dV +\oint\limits_{\partial {\Omega}}
 {{\delta^{\vee} F}\over {\delta q}}\cdot \delta q \vert _{\partial
{\Omega}} dS,\label{eq:lmmr1}
\end{equation}
\begin{equation}
D_{q}F(q,p)\cdot \delta
p=\int\limits_{{\Omega}}{{\delta^{\wedge}F}\over{\delta p}} \cdot \delta
p\ dV +\oint\limits_{\partial {\Omega}}
 {{\delta^{\vee} F}\over {\delta p}}\cdot \delta p \vert _{\partial
{\Omega}} dS.\label{eq:lmmr1a}
\end{equation}
The idea is to generalize
the definition of the variational derivative by incorporating the boundary
contribution
\begin{equation}
\frac{\delta F}{\delta
q}={{\delta^{\wedge}F}\over{\delta q}}+ \delta(S)\cdot{{\delta^{\vee}
F}\over {\delta q}}\label{eq:lmmr_varder}.
\end{equation}
\begin{equation}
\frac{\delta F}{\delta p}={{\delta^{\wedge}F}\over{\delta
p}}+ \delta(S)\cdot{{\delta^{\vee} F}\over {\delta
p}}\label{eq:lmmr_varder1}.
\end{equation}
Then LMMR in fact proposed to
use for the new Poisson bracket the old formula
\[
\{F,G\}=\int\limits_\Omega\Biggl[ {{\delta F}\over {\delta q(x)}}{{\delta
G} \over {\delta p(x)}}- {{\delta G}\over {\delta q(x)}}{{\delta F}\over
{\delta p(x)}} \Biggr] dV
\]
but with  new variational derivatives
(\ref{eq:lmmr_varder}), (\ref{eq:lmmr_varder1}) in it
\[
 \{F,G\}=\int\limits_{{\Omega}} \Biggl[ {{\delta^{\wedge} F}\over {\delta
q(x)}}{{\delta^{\wedge} G} \over {\delta p(x)}}- {{\delta^{\wedge} G}\over
{\delta q(x)}}{{\delta^{\wedge} F}\over {\delta p(x)}} \Biggr] dV \] \[
+\oint\limits_{\partial {\Omega}} \Biggl[ {{\delta^{\wedge} F}\over
{\delta q(x)}}\Bigg\vert _{\partial {\Omega}}{{\delta^{\vee} G} \over
{\delta p(x)}}+
 {{\delta^{\vee} F}\over {\delta q(x)}}{{\delta^{\wedge} G} \over
 {\delta p(x)}}\Bigg\vert _{\partial {\Omega}}\Biggr] dS \] \[
 -\oint\limits_{\partial {\Omega}} \Biggl[ {{\delta^{\wedge} G}\over
{\delta q(x)}}\Bigg\vert _{\partial {\Omega}}{{\delta^{\vee} F} \over
{\delta p(x)}}+ {{\delta^{\vee} G}\over {\delta q(x)}}{{\delta^{\wedge} F}
\over {\delta p(x)}}\Bigg\vert _{\partial {\Omega}}\Biggr] dS.
\]
One immediately sees that the most dangerous term with the product
of $\delta$-functions is absent above. In fact, to kill this term LMMR put
a special boundary condition
\begin{equation}
{{\delta^{\vee}
F}\over{\delta q}}{{\delta^{\vee} G}\over{\delta p}}- {{\delta^{\vee}
G}\over{\delta q}}{{\delta^{\vee} F}\over{\delta p}} =0, \label{eq:lmmr2}
\end{equation}
which enforces zero value for the coefficient standing before this
dangerous product.\footnote{It seems interesting to mention that
in recent calculations of the central charges arising in boundary
symmetry algebras in Chern-Simons theory \cite{park} the similar
term is cancelled automatically due to the Poisson structure
itself.} Unfortunately it is not quite clear whether the Poisson
bracket $\{F,G\}$ preserves this property in general case even if
the initial functionals $F$ and $G$ satisfy (\ref{eq:lmmr2}).

Here we see the bifurcation point for the following generalizations of
LMMR result. The idea of \cite{Bering} is that these dangerous terms with
products of $\delta$-functions must be omitted independently of any
boundary conditions (``annihilation principle'').  Another idea advocated
earlier in \cite{Sol93} is to find a reasonable formula for these
terms.

To explain this in more detail we need first to introduce  a relevant
formalism for treating  general variations of functionals depending on
arbitrary (but finite!) number of spatial derivatives.  In \cite{Sol93}
the adequate
mathematical machinery
  was found to be the so-called higher Eulerian operators
\cite{olv,and,ald}. We shall follow notations of \cite{Sol93,Sol96,Sol98}.
Einstein rule is used, i.e. we omit the sign of summation over repeated
indices and multi-indices.

The first variation of a general local
(${\rm max}|J|<\infty$) functional
\[
F=\int\limits_\Omega f(\phi_A(x),\phi^{(J)}_A(x))d^nx
\]
can be written in a form
\begin{equation}
\delta F= %\sum_A\sum_J
\int\limits_\Omega
\frac{\partial f}{\partial\phi_A^{(J)}}D_J\delta\phi_Ad^nx\equiv
\int\limits_\Omega f'_A(\delta\phi_A)d^nx \equiv %\sum_A\sum_J
\int\limits_\Omega D_J\left(E^J_A(f)\delta\phi_A\right)d^nx,\label{eq:fullvar}
\end{equation}
where in general $J$
denotes multi-index $J=(j_1,...,j_n)$ and
\[
\phi^{(J)}_A=\frac{\partial^{|J|}\phi_A}{\partial^{j_1}x^1
\ldots\partial^{j_n}x^n}\equiv D_J\phi_A,\qquad |J|=j_1+...+j_n.
\] but
in the simplest case of one-dimensional space it is just the order of
spatial
derivative.  We have introduced also the Fr\'echet derivative which is a
differential operator
\begin{equation}
f'_A=\frac{\partial f}{\partial\phi^{(J)}_A}D_J.\label{eq:fr}
\end{equation}
The higher Eulerian
operators $E^J_A$ are uniquely defined by the following formula
\begin{equation} E^J_A(f)=
%\sum_K
{K\choose J}(-D)_{K-J}{{\partial f}
\over{\partial\phi_A^{(K)}}}.\label{eq:he}
\end{equation}
Here
binomial coefficients for multi-indices are
\[
{J \choose K}={j_1\choose k_1}\cdots{j_n\choose k_n},
\]
\[
{j \choose k}= \cases {
j!/(k!(j-k)!) & if  $ 0\le k \le j$; \cr
0  & otherwise, \cr }
\]
and
\[
(-D)_K=(-1)^{|K|}D_K.
\]
Both
\cite{Bering} and \cite{Sol93} use the full variation
(\ref{eq:fullvar}) for
the
construction of the new Poisson brackets but in different ways. In
\cite{Sol93} it was proposed to start from the formula
\[
\{F,G\}=\delta_G F\equiv
%\sum_A\sum_J
\int\limits_\Omega D_J\bigl(E^J_A(f)\delta_G\phi_A\bigr)d^nx\equiv
\int\limits_\Omega f'_A(\delta_G\phi_A)d^nx,
\]
and to look for
$\delta_G\phi_A$ of such a form which fulfils the equation
\[
\delta_G F=-\delta_F G.
\]
 The following formula was derived  in
\cite{Sol93} after some calculations
\[
\{F,G\}= %\sum_{A,B}\sum_{J,K}
\int\limits_\Omega D_{J+K}\left(
E^J_A(f)\hat I_{AB}E^K_B(g)\right)d^nx\equiv \int\limits_\Omega \Tr
\left(f'_A \hat I_{AB}g'_B\right)d^nx. \label{eq:solo}
\]
We can for the easier comparison with \cite{Bering} first consider only
the so-called
ultralocal brackets then
\[
\{\phi_A(x),\phi_B(y)\}=I_{AB}\delta(x,y),\quad I_{AB}=-I_{BA}.
\]
In
contrast, the proposal of \cite{Bering} is to start with the already
antisymmetric expression
\[
\{F,G\}=\Delta_G F-\Delta_F G -\{F,G\}_{\rm
old}, \] where \[ \{F,G\}_{\rm old}= %\sum_{A,B}
\int\limits_\Omega
E^0_A(f)I_{AB}E^0_B(g)d^nx,
\]
\[
\Delta_G F=
%\sum_A\sum_J
\int\limits_\Omega D_J(E^J_A(f)\Delta_G\phi_A)d^nx\equiv
\int\limits_\Omega f'_A(\Delta_G\phi_A)d^nx.
\]
Then it is possible to
use the standard expression for the field variation
\[
\Delta_G\phi_A=I_{AB}E^0_B(g)
\]
and the resulting formula will be
\[
\{F,G\}=
%\sum_{A,B}\sum_J
\int\limits_\Omega D_J\Bigl(E^J_A(f)I_{AB}E^0_B(g)
-E^J_A(g)I_{AB}E^0_B(f)\Bigr)d^nx-\{F,G\}_{\rm old}.  \label{eq:ber}
\]

So, it is easy to see that the last formula contains only one summation
over multi-index $J$ whereas formula (\ref{eq:solo}) contains a double sum
over $J$ and $K$. If we omit all the terms without at least one of $E^0$
operators in this double sum we immediately get (\ref{eq:ber}).

Maybe it will be of some interest to add that in ultralocal case
for the local functionals depending on the spatial derivatives of
the fields of order up to $N$, Bering's
bracket involves spatial derivatives of order $3N$, whereas the bracket
proposed in \cite{Sol93} involves $2N$, as also the standard bracket does.

The point of difficulty with Bering's formula seems to be the lack of
invariance under the changes of dependent variables (differential
substitutions of fields).

\section{Differential substitutions}
Let us consider the invariance properties of the field theory Poisson
brackets under field redefinitions of the type
\begin{equation}
\phi_A \rightarrow \bar\phi_{\bar B}=\xi_{\bar
B}(\phi_A,D_J\phi_A),\label{eq:0}
\end{equation}
(differential substitutions).

If we initially have some local Poisson brackets for fields $\phi_A(x)$,
i.e.
\begin{equation}
\{\phi_A(x),\phi_B(y)\}=\hat I_{AB}(x)\delta(x,y),\label{eq:1}
\end{equation}
where
$
\hat I_{AB}=I^K_{AB}D_K
$
is a differential operator of a finite order with field-dependent
coefficients
\begin{equation}
I^K_{AB}=I^K_{AB}(\phi_C,D_J\phi_C),\label{eq:2}
\end{equation}
then as a result of the differential substitution (\ref{eq:0}) we get a result
\[
\{{\bar\phi}_{\bar C}(x),{\bar\phi}_{\bar D}(y)\}=\left(\xi_{\bar C}
\right)'_A(x)\left(\xi_{\bar D}
\right)'_B(y)\hat I_{AB}(x)\delta(x,y).
\]
To transform this expression to the form similar to (\ref{eq:1}) we need a
definition of the ``adjoint'' operator
\[
\hat J_{AB}(x)\delta(x,y)={\hat J}^{\rm ``adjoint''}_{AB}(y)\delta(x,y),
\]
then we will have
\[
\left(\xi_{\bar C}
\right)'_A(x)\left(\xi_{\bar D}
\right)'_B(y)\hat I_{AB}(x)\delta(x,y)=\left(\xi_{\bar C}
\right)'_A(x)\hat I_{AB}(x)\left[\left(\xi_{\bar D}
\right)'_B\right]^{\rm ``adjoint''}(x)\delta(x,y),
\]
and
\[
\hat I_{\bar C\bar D}(x)=\left(\xi_{\bar C}
\right)'_A(x)\hat I_{AB}(x)\left[\left(\xi_{\bar D}
\right)'_B\right]^{\rm ``adjoint''}(x).
\]
The approach which we consider here is different from the standard one
by
the treatment of boundary (or divergence) terms. All of them should be
preserved in the calculations. This means that we require the exact
equality
\[
\int\limits_{\Omega}
\xi_A\hat J_{AB}\eta_B
d^nx=\int\limits_{\Omega}
\eta_A\hat {J^\dagger}_{AB}\xi_B d^nx,
\]
without discarding any boundary (divergence) terms. In contrast
the standard approach requires only equality up to boundary terms
\[
\xi_A\hat J_{AB}\eta_B\approx\eta_A\hat {J^\ast}_{AB}\xi_B \quad ({\rm
mod \ divergences}),
\]
or, in notations of Appendix C,
\[
\langle\xi|\hat J|\eta\rangle=\langle\eta|\hat J^\ast|\xi\rangle.
\]
From the last definition we get a usual relation
\begin{equation}
{\hat J}^\ast_{AB}=(-D)_K\circ J^K_{BA}.\label{eq:3}
\end{equation}
But  the former one gives a different result:
\begin{equation}
\hat {J^\dagger}_{AB}=(-D)_K\circ
\theta_{\Omega}J^K_{BA}.\label{eq:3a}
\end{equation}
Here we use the characteristic function of the domain of integration
(physical domain)
\[
\theta_{\Omega}(x)= \cases {
1 &  if $x\in\Omega$; \cr
0  & otherwise; \cr }
\]
to  codify the divergences. There is an apparent relation
\[
\int\limits_{\Rn}(-D)_K\theta_{\Omega}f(x)d^nx=\int\limits_{\Rn}\theta_{\Omega}D_Kf(x)d^nx\equiv
\int\limits_{\Omega}D_Kf(x)d^nx.
\]
So, with $\theta_\Omega$ we are able to write all the spatial integrals
not as integrals over
the physical domain but as integrals over the whole infinite coordinate
space $\Rn$.

\section{The standard bracket}
Let us first consider transformations  of the standard Poisson bracket
\begin{equation}
\{F,G\}_{\rm old}=\frac{1}{2}\int\limits_{\Omega}\left(E^0_A(f)\hat
I_{AB}E^0_{B}(g)- E^0_A(g)\hat
I_{AB}E^0_{B}(f)\right)d^nx,
\label{eq:4}
\end{equation}
under field redefinitions of the type (\ref{eq:0}). We will use a formula
\begin{eqnarray}
E^0_{A}(f)&=&\left[\left(\xi_{\bar C} \right)'_{A} \right]^\ast
E^0_{\bar C}(f)\equiv\nonumber\\
&\equiv&\left(E^0_{A}(\xi_{\bar C})-E^1_{A}(\xi_{\bar C})D
+E^2_{A}(\xi_{\bar C})D_2-\ldots \right)E^0_{\bar C}(f)\equiv\nonumber\\
&\equiv& (-1)^{|K|}E^K_{A}(\xi_{\bar C})D_K
E^0_{\bar C}(f),
\end{eqnarray}
derived in Appendix C. Then in the integrand of (\ref{eq:4}) we will have
the following expressions
\begin{eqnarray}
E^0_{A}(f)\hat I_{AB}E^0_{B}(g)&=&\left[\left(\xi_{\bar C}
\right)'_{A} \right]^\ast E^0_{\bar C}(f)\hat I_{AB}\left[\left(\xi_{\bar
D}
\right)'_{B} \right]^\ast E^0_{\bar D}(g)=\nonumber \\
&=&E^0_{\bar C}(f)\left[\left[\left(\xi_{\bar C}
\right)'_{A} \right]^\ast\right]^\dagger \hat I_{AB}\left[\left(\xi_{\bar
D}
\right)'_{B} \right]^\ast E^0_{\bar D}(g).\label{eq:5}
\end{eqnarray}
But, as we discussed above, the two definitions of the ``adjoint''
differential
operator are not equivalent
\[
\left[\left[\left(\xi_{\bar C}
\right)'_{A} \right]^\ast\right]^\dagger=\left(\xi_{\bar C}
\right)'_{A}\circ\theta_\Omega \neq
\left(\xi_{\bar C}
\right)'_{A},
\]
and so
\[
\left[\left[\left(\xi_{\bar C}
\right)'_{A} \right]^\ast\right]^\dagger \hat I_{AB}\left[\left(\xi_{\bar
D}
\right)'_{B} \right]^\ast \neq \hat I_{\bar C \bar D}.
\]
Of course, the difference has a form of  divergences.
Therefore we  see that the standard field theory Poisson bracket is
invariant under field redefinitions of the form (\ref{eq:0}) (i.e.
differential substitutions) only up to boundary terms. Surely, this is
enough because all the other requirements (Jacobi identity, antisymmetry,
Leibnitz rule) are also fulfilled only up to boundary terms and this
bracket does not pretend to be adequate for nontrivial boundary problems.

\section{The general approach to both  proposals \cite{Bering} and
\cite{Sol93}}
Dealing with local functionals and local Poisson brackets in field theory
we always get a general bracket of two functionals in the following  form
\[
\{F,G\}=\int\limits_{\Omega}d^nx\int\limits_{\Omega}d^ny
f'_A(x)g'_B(y)\{\phi_A(x),\phi_B(y)\},
\]
 where Fre\'chet derivatives $f'_A(x)$, $g'_B(y)$ are differential
operators (\ref{eq:fr}). If we represent each
integral not as the integral over finite
domain but as an integral over all the infinite $\Rn$ with a
characteristic function of
the domain $\Omega$
\[
\{F,G\}=\int\limits_{\Rn}\theta_{\Omega}(x)d^nx\int\limits_{\Rn}
\theta_{\Omega}(y)d^ny
f'_A(x)g'_B(y)\{\phi_A(x),\phi_B(y)\},
\]
then it easy to integrate by
parts formally and  get
\[
\{F,G\}=\int\limits_{\Rn} d^nx\int\limits_{\Rn} d^ny
\frac{\delta
F}{\delta\phi_A(x)}\frac{\delta
G}{\delta\phi_B(y)}\{\phi_A(x),\phi_B(y)\},
\]
where
\[
\frac{\delta F}{\delta\phi_A(x)}=\left(f'_A(x)
\right)^\dagger\theta_{\Omega}=
(-D)_K\left(\theta_{\Omega}\frac{\partial f}{\partial\phi_A^{(K)}}\right)
\equiv E^J_A(f)(-D)_J\theta_{\small \Omega},
\]
with the the analogous formula  valid for the variational
derivative of $G$.
Let us restrict the consideration to ultralocal case for simplicity,
then
\[
\{F,G\}=\int\limits_{\Rn} d^nx\int\limits_{\Rn} d^ny
\frac{\delta
F}{\delta\phi_A(x)}I_{AB}\frac{\delta
G}{\delta\phi_B(y)}\delta(x,y),
\]
and in both approaches it is believed that
\[
\{F,G\}=\int\limits_{\Rn} d^nx \frac{\delta
F}{\delta\phi_A(x)}I_{AB}\frac{\delta
G}{\delta\phi_B(x)}.
\]
But, of course, this expression includes products of distributions of the
form
$D_J\theta_\Omega\times D_K\theta_\Omega$ and here these two proposals
are different:
\begin{itemize}
\item in Bering's approach \cite{Bering}:
\[
D_J\theta_\Omega\times
D_K\theta_\Omega=\delta_{J0}D_K\theta_\Omega+\delta_{K0}D_J\theta_\Omega
-\delta_{J0}\delta_{K0}\theta_\Omega,
\]
\item in the approach of \cite{Sol93}:
\[
D_J\theta_\Omega\times D_K\theta_\Omega=D_{J+K}\theta_\Omega.
\]
\end{itemize}
Apparently it is possible to avoid these distributions completely as they
serve only to codify divergences. This is demonstrated in publications
\cite{Sol98}. Then the key transformation from the double spatial integral
to the single spatial integral with the help of $\delta$-function can be
interpreted simply as a pairing between 1-forms and 1-vectors of the
formal variational calculus \cite{olv,GD,Dorf}.
The pairing defined in \cite{Sol93,Sol98} is compatible with the grading
related to divergences.

Now if we use the above formula for Bering's pairing it is possible to derive
the Poisson brackets in the most general (not treated in
\cite{Bering})  non-ultralocal case where $\hat I_{AB}$ is a differential
operator of a finite order with field dependent coefficients.
Really, this is the same formula with hats added\footnote{It is possible
to derive
a more general formula if we do not suggest that the operator $\hat I_{AB}$ should
 be antisymmetric with respect to the standard definition of the adjoint.
But this formula does not fulfil the Jacobi identity even for non-ultralocal
brackets with constant coefficients.}
\begin{eqnarray}
\{f,g\}&=&f'_A\left(\hat I_{AB}E^0_B(g)\right)-g'_A\left(\hat
I_{AB}E^0_B(f)\right)-\nonumber\\
&-&\frac{1}{2}\left(E^0_A(f)\hat I_{AB}E^0_B(g)-
E^0_A(g)\hat I_{AB}E^0_B(f)\right).\nonumber
\end{eqnarray}
Moreover, it is possible to demonstrate that the Jacobi identity is
fulfilled for this bracket for any local operator $\hat I_{AB}$ with
constant coefficients . In the  case of ultralocal Poisson brackets
with the
field dependent
coefficients (but the dependence on field derivatives is excluded)
in Appendix B we derive the following condition of the
fulfilment of the
Jacobi identity for the bracket constructed accroding to Bering proposal
\[
I_{AB,C}I_{CD}+I_{DA,C}I_{CB}+I_{BD,C}I_{CA}=0.
\]
In Appendix B we also give the condition for the most general case
of non-ultralocal brackets with coefficients depending
 on  the spatial derivatives of fields also.

\section{Differential substitutions and Bering's proposal \cite{Bering}}
Now let us consider the  formula derived in the previous Section
as a further development of the initial proposal by Bering \cite{Bering}
\begin{equation}
\{F,G\}_B=\int\limits_{\Omega}f'_A \hat I_{AB}E^0_B(g)d^nx-
\int\limits_{\Omega}g'_A \hat
I_{AB}E^0_B(f)d^nx-\{F,G\}_{\rm old},\label{eq:6}
\end{equation}
where $\{F,G\}_{\rm old}$ is the standard Poisson bracket treated before.

The Fr\'echet derivative transforms under differential substitutions
(\ref{eq:0}) as follows (see Appendix C)
\[
f'_A=f'_{\bar C}\left(\xi_{\bar C} \right)'_A,
\]
 so we obtain
\[
f'_A\hat I_{AB}E^0_B=f'_{\bar C}\left(\xi_{\bar C} \right)'_{A}\hat
I_{AB}\left[\left(\xi_{\bar D} \right)'_B \right]^\ast E^0_{\bar D}(g).
\]
It means that the first and second terms of the bracket will be
invariant if we suppose
\[
\hat I_{\bar C\bar D}=\left(\xi_{\bar C} \right)'_A\hat I_{AB}\left[\left(
\xi_{\bar D}\right)'_B
\right]^\ast,
\]
so, the old definition of the adjoint operator (\ref{eq:3}) should be used
here in accordance with the treatment given by Bering (see Subsection
(5.5) of
\cite{Bering}).

Unfortunately, this bracket contains also
a term $\{F,G\}_{\rm old}$ (the standard Poisson bracket) with another
transformation properties. As we have
demonstrated in  Section 3 this term is invariant under
field redefinitions only up to
divergences. So, taken as a whole, Bering's formula is not invariant.

\section{Differential substitutions and the bracket proposed in
\cite{Sol93}}
Let us show that in contrast the formula
\[
\{F,G\}=\int\limits_{\Omega}\Tr\left(f'_A\hat I_{AB}g'_B \right)d^nx
\]
is precisely invariant under field redefinitions of the form (\ref{eq:0}).
We remind that the trace is used here to denote the rules of composition
of the differential operators $f'_A$, $\hat I_{AB}$ and $g'_B$:
\begin{eqnarray}
f'_A&=&\frac{\partial f}{\partial\phi^{(J)}_A}D_J,\nonumber\\
\hat I_{AB}&=&I^K_{AB}D_K,\nonumber\\
g'_B&=&\frac{\partial g}{\partial\phi^{(J)}_B}D_J.\nonumber
\end{eqnarray}
Operator $f'_A$ acts on everything to the right of it, so does operator
$\hat I_{AB}$, and operator  $g'_B$ acts on everything to the left of it,
i.e. acts on everything besides its own coefficients,
\[
\Tr\left(f'_A\hat I_{AB}g'_B
\right)\equiv{J \choose L}{K \choose M}D_M\frac{\partial
f}{\partial\phi^{(J)}_A}D_{J+K-L-M}\hat I_{AB}D_L\frac{\partial
g}{\partial\phi^{(K)}_B}.
\]
After the field redefinition we get
\[
\{F,G\}=\int\limits_{\Omega}\Tr\left(f'_{\bar C}(\xi_{\bar C})'_A\hat
I_{AB}g'_{\bar D}(\xi_{\bar D})'_B \right)d^nx.
\]
So, if we use here the adjoint operator to $(\xi_{\bar D})'_B $
defined by
(\ref{eq:3a})
then it will act only onto $g'_{\bar D}$
\[
\{F,G\}=\int\limits_{\Omega}\Tr\left(f'_{\bar C}(\xi_{\bar C})'_A\hat
I_{AB}\left[(\xi_{\bar D})'_B\right]^\dagger g'_{\bar D} \right)d^nx.
\]
But according to our definitions given in Section 2
\[
\hat I_{\bar C\bar D}=(\xi_{\bar C})'_A\hat I_{AB}\left[(\xi_{\bar
D})'_B\right]^\dagger .
\]
As a result we see that this definition of the field theory Poisson
bracket with boundary terms is exactly invariant under differential
substitutions.
%field redefinitions.

In publication \cite{Sol97} this invariance was demonstrated for the
concrete example --- the Ashtekar transformation \cite{Ash} of the
gravitational variables.

\section{Conclusion} We considered above an interesting proposal made by
Bering on the boundary terms in the field theory Poisson bracket.
We generalize this proposal to the most general local Poisson brackets and find
the conditions necessary to fulfil the Jacobi identity.
According to our treatment given in more detail in previous publications
\cite{Sol93,Sol98} there are three separate ingredients of the Poisson
bracket construction which should be revised: the differential of the
local functional, the Poisson bivector and the pairing operation.
Bering uses the same definition for the differential, but changes
the pairing and the bivector. It occurs so that to change the
pairing alone means to get into trouble with the Jacobi identity
in the non-ultralocal case.

Really,
the paper \cite{Bering} suggest a lot of new ideas which deserve more
discussion. Here we only concentrated on the  drawback  that it
seemingly had. Probably, the further investigation will show whether these
drawback could be overcome in Bering's approach. But anyhow it is
absent if we use another formula suggested in \cite{Sol93}.

\section*{Acknowledgements}
This work was completed during the visit of the author to the
International Center for Theoretical Physics. The author is most grateful
to ICTP for hospitality and to Professor
S.~Randjbar-Daemi for the invitation.

\section*{Appendix A. Useful relations} Let us suppose that $\hat
I_{AB}=I^M_{AB}D_M$, where $I^M_{AB}=I^M_{AB}(\phi^{(J)})$, and
\[
{\hat I_{AB}}^\ast=(-D)_M\circ I^M_{BA}.
\]
Then the following
useful relations can be proved by using the technique of higher Eulerian
operators \cite{olv,ald} compiled in Lemmas 2.5 -- 2.12 of \cite{Sol93}:
\[
\left(E^0_B(g)\right)'_C(\hat I_{CD}E^0_D(h))=
(-D)_L\left[\frac{\partial^2
g}{\partial\phi^{(L)}_B\partial\phi^{(J)}_C}D_J(\hat I_{CD}E^0_D(h))
\right],
\]
\begin{eqnarray}
E^0_B\left(g'_C(\hat I_{CD}E^0_D(h))
\right)&=&
 (-D)_L\left[\frac{\partial^2
g}{\partial\phi^{(L)}_B\partial\phi^{(J)}_C}D_J(\hat
I_{CD}E^0_D(h))+\right.\nonumber\\
&+&\left.\frac{\partial^2
h}{\partial\phi^{(L)}_B\partial\phi^{(J)}_C}D_J({\hat I_{CD}}^\ast
E^0_D(g))+\right.\nonumber\\
&+&\left. E^0_C(g)\frac{\partial
I^M_{CD}}{\partial\phi^{(L)}_B}D_M E^0_D(h)  \right],\nonumber\\
E^0_B\left(E^0_C(g)\hat I_{CD}E^0_D(h) \right)&=&
(-D)_L\left[\frac{\partial^2
g}{\partial\phi^{(L)}_B\partial\phi^{(J)}_C}D_J(\hat
I_{CD}E^0_D(h))+\right.\nonumber\\ &+&\left.\frac{\partial^2
h}{\partial\phi^{(L)}_B\partial\phi^{(J)}_C}D_J({\hat I_{CD}}^\ast
E^0_D(g))\right.\nonumber\\ &+& \left. E^0_C(g)\frac{\partial
I^M_{CD}}{\partial\phi^{(L)}_B}D_M E^0_D(h)  \right],\nonumber
\end{eqnarray}

Let also ${\hat I}^\ast_{AB}=-\hat I_{AB}$, then
from the above it follows
\begin{eqnarray}
E^0_B\left(g'_C(\hat I_{CD}E^0_D(h)) \right)=-E^0_B\left( h'_C(\hat
I_{CD}E^0_D(g))\right)=\nonumber\\
=E^0_B\left(E^0_C(g)\hat I_{CD}E^0_D(h)
\right)
=(-D)_L\left[E^0_C(g)\frac{\partial
I^M_{CD}}{\partial\phi^{(L)}_B}D_M E^0_D(h)  \right]+\nonumber\\
+\left(E^0_B(g)
\right)'_C(I_{CD}E^0_D(h))-\left(E^0_B(h)
\right)'_C(I_{CD}E^0_D(g)).\nonumber
\end{eqnarray}
We will use these
relations in checking the Jacobi identity in Appendix B.

Let us illustrate our results by the less general case of ultralocal
Poisson brackets
\[
\hat I_{AB}=I_{AB}=-I_{BA},
\]
and  suppose for the simpllicity that functions  $I_{AB}$
depend only on fields and do not depend on their derivatives
\[
I_{AB}=I_{AB}(\phi_C).
\]
Then
\[
\frac{\partial
I^M_{CD}}{\partial\phi^{(L)}_B}=\delta_{M0}\delta_{L0}I_{CD,B},
\]
and so,
\[
(-D)_L\left[E^0_C(g)\frac{\partial
I^M_{CD}}{\partial\phi^{(L)}_B}D_M E^0_D(g)  \right]=I_{CD,B}E^0_C(g)
E^0_D(h).
\]

\section*{Appendix B. The Jacobi identity} By using Bering's formula for
the Poisson bracket
we get \begin{eqnarray}
\{F,G\}&=&\int\limits_{\Rn}\{f,g\}d^nx,\nonumber\\
\{f,g\}&=&f'_A(\hat I_{AB}E^0_B(g))-g'_A(\hat
I_{AB}E^0_B(f))-\nonumber\\ &-&\frac{1}{2}E^0_A(f)\hat
I_{AB}E^0_B(g)+\frac{1}{2}E^0_A(g)\hat I_{AB}E^0_B(f),\nonumber\\
\{\{f,g\},h\}&=& { \{f,g\} }'_C (\hat I_{CD}E^0_D(h))-h'_C( \hat
I_{CD}E^0_D(\{f,g\}))- \nonumber\\ &-&\frac{1}{2}E^0_C(\{f,g\})\hat
I_{CD}E^0_D(h)+\frac{1}{2}E^0_C(h)\hat I_{CD}E^0_D(\{f,g\})= \nonumber\\
&=&f''_{AC}\left(\hat I_{AB}E^0_B(g),\hat
I_{CD}E^0_D(h)\right)-g''_{AC}\left(\hat I_{AB}E^0_B(f),\hat
I_{CD}E^0_D(h)\right)+\nonumber\\
&+&f'_A\left((\hat I_{AB})'_C(\hat
I_{CD}E^0_D(h))E^0_B(g)+ \hat I_{AB}(E^0_B(g))'_C(\hat
I_{CD}E^0_D(h))\right)-\nonumber \\
&-&g'_A\left((\hat I_{AB})'_C(\hat
I_{CD}E^0_D(h))E^0_B(f)+ \hat I_{AB}(E^0_B(f))'_C (\hat
I_{CD}E^0_D(h))\right)-\nonumber\\
&-&\frac{1}{2}(E^0_A(f))'_C(\hat I_{CD}E^0_D(h))\hat
I_{AB}E^0_B(g)-\nonumber\\
&-&\frac{1}{2}E^0_A(f)(\hat I_{AB})'_C(\hat
I_{CD}E^0_D(h))E^0_B(g)-\nonumber\\
&-&\frac{1}{2}E^0_A(f)\hat I_{AB}(E^0_B(g))'_C(\hat
I_{CD}E^0_D(h))+\nonumber\\
&+&\frac{1}{2}(E^0_A(g))'_C(\hat I_{CD}E^0_D(h))\hat
I_{AB}E^0_B(f)+\nonumber\\
&+&\frac{1}{2}E^0_A(g)(\hat I_{AB})'_C(\hat
I_{CD}E^0_D(h))E^0_B(f)+\nonumber\\
&+&\frac{1}{2}E^0_A(g)\hat I_{AB}(E^0_B(f))'_C(\hat
I_{CD}E^0_D(h))+\nonumber\\
&+&h'_C\left(\hat I_{CD}E^0_D\bigl(g'_A(\hat
I_{AB}E^0_B(f))-f'_A( \hat I_{AB}E^0_B(g))+\right.\nonumber\\
&+&\left.
E^0_A(f)\hat I_{AB}E^0_B(g)) \bigr)\right)-\nonumber\\
&-&\frac{1}{2}E^0_C\left(f'_A(\hat I_{AB}E^0_B(g))-g'_A(\hat
I_{AB}E^0_B(f))-\right.\nonumber\\
&-&\left.\frac{1}{2}E^0_A(f)\hat I_{AB}E^0_B(g)+\frac{1}{2}E^0_A(g)
\hat I_{AB}E^0_B(f)\right)\hat I_{CD}E^0_D(h)+\nonumber\\
&+&\frac{1}{2}E^0_C(h)\hat I_{CD}E^0_D\left(f'_A(\hat
I_{AB}E^0_B(g))-g'_A(\hat
I_{AB}E^0_B(f))-\right.\nonumber\\
&-&\left.\frac{1}{2}E^0_A(f)\hat I_{AB}E^0_B(g)+\frac{1}{2}E^0_A(g)
\hat I_{AB}E^0_B(f)
\right)\nonumber
\end{eqnarray}
Here we use notation \[
f''_{AB}(\xi_A,\eta_B)=\frac{\partial^2 f}{\partial\phi_A^{(J)}
\partial\phi_B^{(K)}}D_J\xi_AD_K\eta_B.  \]
Then by making cyclic
permutations and applying formulae from Appendix A we get a result
\[
\{\{f,g\},h\}+\{\{h,f\},g\}+\{\{g,h\},f\}=
\]
\[
= f'_A\left(\frac{\partial I^M_{AB}}{\partial\phi^{(L)}_C}D_L(\hat
I_{CD}E^0_D(h)D_ME^0_B(g)) \right)
\]
\[
-g'_A\left(\frac{\partial I^M_{AB}}{\partial\phi^{(L)}_C}D_L(\hat
I_{CD}E^0_D(h)D_ME^0_B(f))  \right)
\]
\[
-h'_C\left(\hat I_{CD}(-D)_L\left[\frac{1}{2}E^0_A(f)\frac{\partial
I^M_{AB}}{\partial\phi_D^{(L)}}D_ME^0_B(g)-\frac{1}{2}E^0_A(g)\frac{\partial
I^M_{AB}}{\partial\phi_D^{(L)}}D_ME^0_B(f) \right]  \right)
\]
\[
-\frac{1}{2}E^0_A(f)\frac{\partial
I^M_{AB}}{\partial\phi^{(L)}_C}D_L(\hat I_{CD}E^0_D(h))D_ME^0_B(g)
\]
\[
+\frac{1}{2}E^0_A(g)\frac{\partial
I^M_{AB}}{\partial\phi^{(L)}_C}D_L(\hat I_{CD}E^0_D(h))D_ME^0_B(f)
\]
\[
-\frac{1}{2}(-D)_L\left[\frac{1}{2}E^0_A(f)\frac{\partial
I^M_{AB}}{\partial\phi^{(L)}_C}D_ME^0_B(g)-
\frac{1}{2}E^0_A(g)\frac{\partial
I^M_{AB}}{\partial\phi^{(L)}_C}D_ME^0_B(f)
\right]\hat I_{CD}E^0_D(h)
\]
\[
+\frac{1}{2}E^0_C(h)\hat
I_{CD}(-D)_L\left[\frac{1}{2}E^0_A(f)\frac{\partial
I^M_{AB}}{\partial\phi^{(L)}_D}D_ME^0_B(g)-
\frac{1}{2}E^0_A(g)\frac{\partial
I^M_{AB}}{\partial\phi^{(L)}_D}D_ME^0_B(f)
\right]
\]
\[
+ {\rm cyclic}\
{\rm permutation}\ {\rm of}\ (f,g,h)  .
\]
From the above expression it is apparent that in the case of constant
coefficients $I^M_{AB}$ the Jacobi identity is satisfied.
It is straightforward to check that in the  case of
ultralocal Poisson brackets with the coefficients depending on the fields
(but not on their spatial derivatives) we get a well-known condition for
the fulfilment of the Jacobi identity
\[
I_{AB,C}I_{CD}+{\rm cyclic}\
{\rm permutation}\ {\rm of}\ (A,B,D)=0.
\]

\section*{Appendix C. The transformation rules}
Here we derive the transformation rules for Euler-Lagrange
and Fr\'echet derivatives under differential substitutions of fields (\ref{eq:0}).

First, let us consider the variation of an arbitrary function of the
fields
\[
\delta f=f'_A\delta\phi_A\equiv \frac{\partial
f}{\partial\phi^{(J)}_A}D_J\delta\phi_A.
\]
If we use the transformed fields
\[
\bar\phi_{\bar B}=\xi_{\bar B}(\phi_A,D_J\phi_A),
\]
then we get
\[
\delta f=f'_{\bar B}\delta\bar\phi_{\bar B}\equiv \frac{\partial
f}{\partial\bar\phi^{(K)}_{\bar B}}D_K\delta\bar\phi_{\bar B},
\]
where
\[
\delta\bar\phi_{\bar B}=(\xi_{\bar B})'_A\delta\phi_A.
\]
Therefore,
\[
f'_A=f'_{\bar B}\circ(\xi_{\bar B})'_A.
\]

Second, let us consider an expression
\[
\langle
1|f'_A|\delta\phi_A\rangle=\langle\delta\phi_A|(f'_A)^\ast|1\rangle\equiv
E^0_A(f)\delta\phi_A,
\]
where the angle brackets denote the standard integrand, defined up to
divergences,
and make a change of variables
\[
\phi_A \rightarrow \bar\phi_{\bar B}=\xi_{\bar
B}(\phi_A,D_J\phi_A),
\]
then
\[
E^0_A(f)\delta\phi_A=\langle 1|f'_{\bar B}\circ (\xi_{\bar
B})'_A|\delta\phi_A|\rangle =\langle 1|f'_{\bar B}|(\xi_{\bar
B})'_A\delta\phi_A\rangle=
 \]
\[
=\langle (\xi_{\bar B})'_A\delta\phi_A|(f'_{\bar B})^\ast |1\rangle=
E^0_{\bar B}(f)(\xi_{\bar B})'_A\delta\phi_A=
\]
\[
=\langle E^0_{\bar B}(f)|(\xi_{\bar B})'_A|\delta\phi_A\rangle=
\langle\delta\phi_A|\left((\xi_{\bar B})'_A\right)^\ast|E^0_{\bar
B}(f)\rangle,
\]
or,
\[
E^0_A(f)=\left((\xi_{\bar B})'_A\right)^\ast E^0_{\bar B}(f).
\]
This result can be checked by more tedious but straightforward
calculation by using formulae for Eulerian operators given in
article \cite{ald}.

\hfill 
\end{document}